# Heart Rate Tracking using Wrist-Type Photoplethysmographic (PPG) Signals during Physical Exercise with Simultaneous Accelerometry

Mahdi Boloursaz, Ehsan Asadi, Mohsen Eskandari, Shahrzad Kiani, *Student Members, IEEE*, and Farokh Marvasti, *Senior Member, IEEE*

*Abstract*—This paper considers the problem of casual heart rate tracking during intensive physical exercise using simultaneous 2 channel photoplethysmographic (PPG) and 3 dimensional (3D) acceleration signals recorded from wrist. This is a challenging problem because the PPG signals recorded from wrist during exercise are contaminated by strong Motion Artifacts (MAs). In this work, a novel algorithm is proposed which consists of two main steps of MA Cancellation and Spectral Analysis. The MA cancellation step cleanses the MA-contaminated PPG signals utilizing the acceleration data and the spectral analysis step estimates a higher resolution spectrum of the signal and selects the spectral peaks corresponding to HR. Experimental results on datasets recorded from 12 subjects during fast running at the peak speed of 15 km/hour showed that the proposed algorithm achieves an average absolute error of 1.25 beat per minute (BPM). These experimental results also confirm that the proposed algorithm keeps high estimation accuracies even in strong MA conditions.

## I. INTRODUCTION

PULSE oximeters facilitate noninvasive continuous monitoring of heart rate (HR) by recording Photoplethysmographic (PPG) signals from skin [1]. A PPG signal is obtained by illuminating skin using a light-emitting diode and detecting the intensity changes in the reflected light. Hence the periodicity of the PPG signal represents HR [2].

However, PPG signals are highly contaminated by artifacts caused by movements of the subject. Such motion artifacts (MAs) strongly interfere with HR especially in fitness applications when the PPG signal is recorded during physical exercise of the subject [3]. The situation gets worse in the case considered in this paper when the PPG signals are recorded from wrist. Such PPG signals experience severe MA due to the loose interface between the pulse oximeter and skin [4].

Many signal processing algorithms proposed so far for MA reduction from PPG signals consider weak MA scenarios in which subjects perform small motions such as finger movements [6], [7], [8] and walking [8]. Moreover, some prior works that consider the scenario of HR monitoring from PPG signals in fitness consider low MA PPG signals recorded from fingertip [8], [9] or ear [10]. Although HR monitoring from wrist-type PPG signals during intensive physical exercise is extremely challenging, but it is of great interest to wearable smart devices such as smart-watches [4].

In order to overcome this challenge, we consider that simultaneous acceleration data is also recorded and available for processing. By using this extra information, we propose an efficient algorithm that accurately tracks HR during high speed (12-15 km/hour) running of the subject by simultaneous processing of the PPG signals taken from two different sensors and the acceleration data along the three axes. The proposed algorithm consists of two main steps of MA cancellation and spectral analysis.

The MA cancellation step decomposes the acceleration signals into periodic MA components using Singular Value Decomposition (SVD). These MA components are then suppressed in the PPG signals by adaptive filtering leaving 2 channel cleansed PPG signals with sparse spectrum.

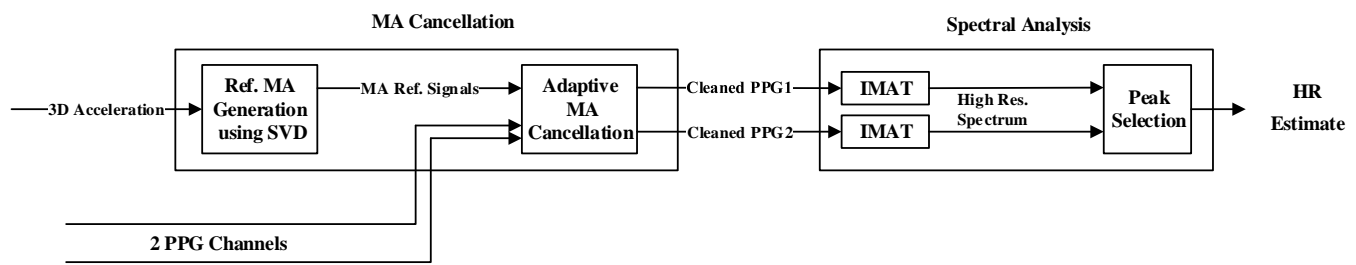

Fig. 1. Block Diagram of the Proposed HR Tracking Algorithm

The Spectral Analysis step applies a few iterations of the IMAT (Iterative Method with Adaptive Thresholding) sparse reconstruction algorithm to achieve a higher resolution spectrum of the cleansed signals and selects the spectral peaks corresponding to HR. These blocks are further explained in section II.

Experimental results on benchmark datasets provided by [4] showed an average absolute error of 1.25 BPM. It should be noted that utilizing the same datasets and performance metrics proposed by [4] facilitates a fair comparison of the achieved results with state-of-the-art techniques [4], [12]-[14] in this field as provided in section III.

"*Contribution:* Although Adaptive Noise Cancelation (ANC) was already utilized by [7]-[8], [15]-[16] to remove motion artifacts from the recorded PPG signals, the reference MA signals were extracted from the PPG signal itself. This approach may work sufficiently well for low artifact scenarios studied in [7]-[8], [15]-[16], but for our fitness scenario, the MA reference signals need to be extracted from simultaneous acceleration data. However, as the acceleration data are convoluted noisy signals composed of different periodic components themselves, it was observed that utilizing them directly as the MA reference (as proposed by [17]) would hamper convergence of the applied adaptive filter and degrade the overall performance. Hence, another key innovation of this research is to extract the reference MA signals by decomposing the 3D acceleration signals by SVD prior to adaptive filtering. The significance of this decomposition substep is observed by simulations in Table 3, section III. This substep also provides an additional design parameter (the number of reference MA signals extracted) to let us trade the estimation error with the simulation time. The proposed *Peak Selection* sub step is another key innovative aspect of this research that efficiently tracks HR by simultaneous processing of the two channel PPG signals in comparison with [4], [11] and [12] that consider single channel PPG signals available. Finally, avoiding the high complexity Multiple Measurement Vector (MMV) model utilized by [12] and utilizing the Iterative Method with Adaptive Thresholding (IMAT) that proved to maintain acceptable spectrum estimation performance while reducing the computational complexity in comparison with FOCUSS [18] used by [4] is a key improvement achieved by this research."

The rest of this paper is organized as follows. Section II explains the proposed method. Section III presents and discusses the simulation results achieved by the proposed method on the 12 benchmark datasets. And finally section IV concludes this paper.

For further reproduction of the reported results, MATLAB codes of the proposed algorithm have been made available online at: ee.sharif.edu/~imat/

## II. THE PROPOSED ALGORITHM

As mentioned earlier, the proposed algorithm consists of two main steps of MA cancellation and spectral analysis. Fig. 1 shows a block diagram of the proposed HR tracking algorithm. The blocks used in the proposed algorithm are explained in the following subsections.

### A. MA Cancellation

In this step, prior to further processing, we filter both PPG signals in [0.4-5] Hz band to reject the MAs outside the natural HR range. Once the out-of-band MAs are cancelled, we apply adaptive filters to suppress in-band MAs. The reference MA signal components needed for the adaptive filter are extracted from the simultaneous acceleration data by Singular Value Decomposition (SVD). These two substeps are further explained in the following:

*Reference MA Generation using SVD:* As expected, the simultaneous acceleration data along the three axes include footprints of the MAs. However, the acceleration data are convoluted noisy signals composed of different periodic components themselves. This will strongly hamper convergence of the applied adaptive filter. Hence, we propose to generate the reference MA signals by decomposing the three acceleration signals by SVD prior to adaptive filtering as depicted in Fig. 1. In this technique, each acceleration signal goes through embedding, SVD and grouping steps [4]. During embedding, a so called L-trajectory matrix is formed from each acceleration signal. Subsequently, these L-trajectory matrices are decomposed to $d$ linearly independent rank-one matrices. These $d$ matrices are classified into $g$ groups with the same or harmonically related oscillatory components. Finally, $n$ groups with frequency components inside [0.4-5] Hz band are utilized as the MA reference signal components for adaptive cancellation.

*Adaptive MA Cancellation:* During this procedure, all the reference MA components derived by SVD are removed from both PPG signals by successive application of adaptive filters as depicted in Fig. 2. Each adaptive filter stage receives the residual signal resulting from its prior stage and a reference signal component as input. These adaptive filters remove both periodic and random MA components. Finally, the resulting cleansed signal is input to the spectral analysis step for HR tracking.

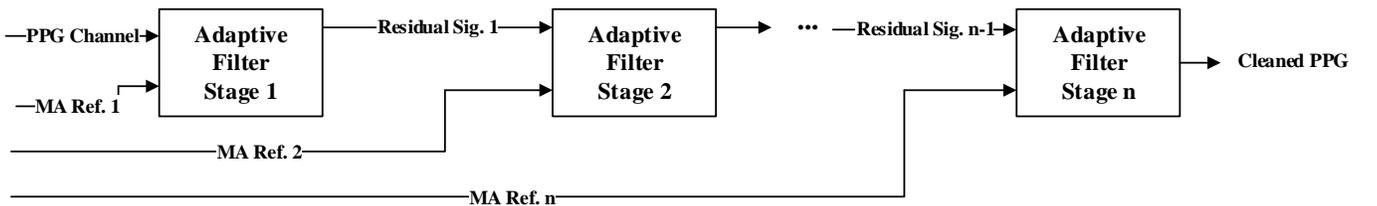

Fig. 2. MA Cancellation by Successive Adaptive Filter Stages

## B. Spectral Analysis

Spectral Analysis is a key step in the proposed HR tracking algorithm. As observed in Fig. 1, this step consists of two substeps of High Resolution Spectrum Estimation using the Iterative Method with Adaptive Thresholding (IMAT) and Peak Selection. IMAT is a fast and efficient algorithm for sparse signal reconstruction [19], [20] that proved to outperform some other spectrum estimation techniques regarding reconstruction performance and complexity [21]. This algorithm is used to provide a higher resolution and denoised spectrum of the input signal prior to Peak Selection. In Peak Selection, we apply some decision mechanisms to accurately estimate the frequency location index of HR in each time window. These decision mechanisms are thoroughly explained below.

*Peak Selection:* The decision mechanisms used in Peak Selection exploit the frequency harmonic relation of HR along with the assumption that HR varies smoothly along time and does not experience jumps between successive time windows. In fact, our experiments showed that in many cases the spectral peak associated with HR keeps its location unchanged in two successive time windows.

To initialize our system, we estimate HR by considering the highest peak in the spectrum of the first 8 second window. This initialization technique is valid as the subjects are required to reduce hand motions during the first few seconds of recording and hence the first PPG window is considered artifact-free.

In the proceeding windows, we track HR by the following algorithm. Denote the estimated HR frequency locations in the current and previous windows by $N_{cur}$ and $N_{prev}$ respectively. Now, consider the three frequency ranges $R_1, R_2, R_3$ in the current time window as in (1):

$$R_1 = [N_{prev} - \epsilon_1, N_{prev} + \epsilon_1]$$
$$R_2 = [2 \times N_{prev} - \epsilon_2, 2 \times N_{prev} + \epsilon_2] \quad (1)$$
$$R_3 = [3 \times N_{prev} - \epsilon_3, 3 \times N_{prev} + \epsilon_3]$$

Denote by $P_1, P_2, P_3$ the frequency location index of the highest peaks in $R_1, R_2, R_3$ and by $S_1, S_2, S_3$ the corresponding signal spectrum values at $P_1, P_2, P_3$. Follow the mentioned steps for both PPG channels and denote by $P_{11}, P_{21}, P_{31}$ the frequency location of the highest peaks in the first PPG channel and $P_{12}, P_{22}, P_{32}$ in the second PPG channel. Now, consider the following three cases.

**Case1:** If one of $S_{11}, S_{21}, S_{31}, S_{12}, S_{22}, S_{32}$ is significantly greater than others (dominant peak) as defined by equation (2), we select the HR frequency index in the current time window as its corresponding fundamental frequency. Note that in (2), $0 < T < 1$ is a predefined threshold optimized by simulations for the best performance.

$$N_{cur} = \frac{P_{ij}}{i} \overset{Case1}{\iff} \forall (k,l) \neq (i,j): S_{ij} \times T > S_{kl} \quad (2)$$

**Case2:** If there is no dominant peak as defined in case 1 in either PPG channels, we search among $P_{11}, P_{21}, P_{31}, P_{12}, P_{22}, P_{32}$ to find a peak pair with a harmonic relation and choose the corresponding fundamental frequency as the HR frequency index in the current time window according to equation (3).

$$N_{cur} = \frac{\frac{P_{ij}}{i} + \frac{P_{kt}}{k}}{2} \overset{Case2}{\iff} \left|\frac{P_{ij}}{i} - \frac{P_{kt}}{k}\right| < \delta \quad (3)$$

**Case3:** If none of the above two cases occur, we make a 10 second time window by concatenating the current and the previous windows and similarly define the three harmonic ranges as above. We also denote the frequency location index of the highest peaks in these ranges by $Q_{11}, Q_{21}, Q_{31}, Q_{12}, Q_{22}, Q_{32}$. Finally, we set $N_{cur}$ as the average of all the available fundamental frequencies as given by equation (4).

$$N_{cur} = \frac{1}{12}(P_{11}, \frac{P_{21}}{2}, \frac{P_{31}}{3}, P_{12}, \frac{P_{22}}{2}, \frac{P_{32}}{3} + Q_{11}, \frac{Q_{21}}{2}, \frac{Q_{31}}{3}, Q_{12}, \frac{Q_{22}}{2}, \frac{Q_{32}}{3}) \quad (4)$$

## III. SIMULATION RESULTS

### A. Parameter Setting

When using SSA, we choose $d = 100$ for the SVD algorithm. An adaptive LMS filter of order 25 was used for MA cancellation with an optimized $\mu = 0.005$. It was also observed that 5 iterations of the IMAT algorithm with optimized parameters $\alpha = 0.1$ yields a high resolution and sufficiently accurate spectrum of the cleansed PPG signals. Finally, the parameters $T$, $\delta$, $\epsilon_1$, $\epsilon_2$ and $\epsilon_3$ are set to 0.6, 9, 60, 80 and 100 respectively.

Table 1. The Results Achieved by the Proposed Algorithm on 12 Subjects' Recordings

|       | Subj1  | Subj2  | Subj3  | Subj4  | Subj5  | Subj6  | Subj7  | Subj8  | Subj9  | Subj10 | Subj11 | Subj12 |
|-------|--------|--------|--------|--------|--------|--------|--------|--------|--------|--------|--------|--------|
| AAE   | 1.72   | 1.33   | 0.90   | 1.28   | 0.93   | 1.41   | 0.61   | 0.88   | 0.59   | 3.78   | 0.85   | 0.71   |
| AEP   | 1.5    | 1.3    | 0.75   | 1.2    | 0.69   | 1.2    | 0.5    | 0.8    | 0.5    | 2.4    | 0.6    | 0.5    |
| EV    | 7.63   | 6.23   | 1.88   | 6.21   | 2.06   | 9.67   | 0.66   | 1.99   | 0.71   | 28.40  | 1.42   | 0.79   |
| PC    | 0.9998 | 0.9998 | 0.9999 | 0.9998 | 1.0000 | 0.9997 | 1.0000 | 0.9999 | 1.0000 | 0.9994 | 1.0000 | 1.0000 |
| ASTPF | 6.36   | 5.94   | 5.93   | 5.92   | 5.94   | 5.93   | 5.93   | 5.93   | 5.93   | 5.96   | 5.92   | 5.91   |

*B. Results*

To observe the effects of motion on the recorded PPG signals of subject 1, the frequency content of a single window of the original and cleansed PPG signals are shown in Fig. 3 (a) and (c), respectively. Fig. 3 (b) also shows the 3D acceleration signals used as the reference for Adaptive MA Cancellation. This figure shows that the MA cancellation step successfully removes motion artifacts from the PPG signals and makes the spectral peak associated with HR more significant.

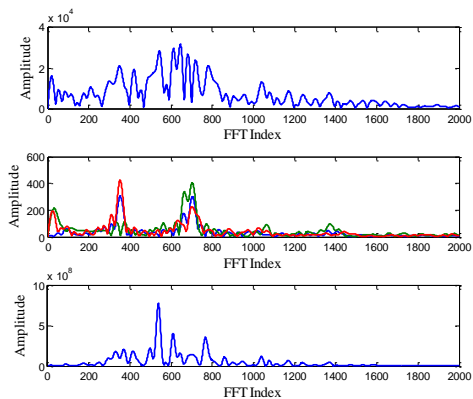

Fig. 3. Frequency Content of a) The Original PPG Signal b) 3D Acceleration Data c) Cleansed PPG Signal

Table 1 lists the achieved Average Absolute Error (AAE), Average Error Percentage (AEP), Estimation Variance (EV), Pearson Correlation (PC) and Average Simulation Time Per Frame (ASTPF) on all 12 subjects' recordings. Note that ASTPF denotes the average amount of time (in seconds) needed on an Intel ® Core ™ i7-4930K CPU @ 3.40 GHz PC to process a single 8 second PPG frame and other reported criteria are defined as by [4]. This results yield an average absolute error of 1.25 BPM.

Table 2. The Results Achieved by the Proposed Algorithm on 12 Subjects' Recordings

| Subject ID | TROIKA [4] | JOSS [12] | SPECTRAP [13] | MICROST [14] | Proposed Method |
|---|---|---|---|---|---|
| 1 | 2.29 | 1.33 | 1.18 | 2.93 | 1.72 |
| 2 | 2.19 | 1.75 | 2.42 | 3.06 | 1.33 |
| 3 | 2.00 | 1.47 | 0.86 | 2.03 | 0.9 |
| 4 | 2.15 | 1.48 | 1.38 | 2.29 | 1.28 |
| 5 | 2.01 | 0.69 | 0.92 | 2.64 | 0.93 |
| 6 | 2.76 | 1.32 | 1.37 | 2.58 | 1.41 |
| 7 | 1.67 | 0.71 | 1.53 | 1.97 | 0.61 |
| 8 | 1.93 | 0.56 | 0.64 | 1.77 | 0.88 |
| 9 | 1.86 | 0.49 | 0.60 | 1.87 | 0.59 |
| 10 | 4.7 | 3.81 | 3.65 | 3.81 | 3.78 |
| 11 | 1.72 | 0.78 | 0.92 | 1.91 | 0.85 |
| 12 | 2.84 | 1.04 | 1.25 | 4.07 | 0.71 |
| Average AAE | 2.34 | 1.29 | 1.50 | 2.58 | 1.25 |

Table 2 compares the performance of the proposed technique with other state-of-the-art techniques that utilize the same datasets and performance measures.

Table 3 reports the AAE values achieved for different number of MA reference signals ($n$) utilized for adaptive MA cancelation. Note that $n = 3$ represents the case where 3D acceleration signals are directly fed to the adaptive filters. It is observed that direct utilization of the acceleration signals as MA reference will hamper the convergence of the adaptive filter and hence degrade the overall performance. This table shows the significance of the proposed reference MA generation algorithm. Overall, we can conclude that $n = 100$ is an optimum choice regarding both performance and complexity. We also observe that the algorithm's simulation time can be decreased by decreasing the design parameter $n$ and tolerating higher AAE values.

Table 3. The AAE values Achieved for different number of MA reference signals on Subject 1 Recordings

| $n$ | 3 | 50 | 100 | 150 | 200 |
|---|---|---|---|---|---|
| Average AAE | 4.83 | 2.45 | 1.72 | 1.98 | 2.18 |
| Average ASTPF | 0.36 | 2.71 | 5.84 | 9.82 | 15.08 |

Finally, table 4 compares the AAE and ASTPF values achieved for Subject 1 while using IMAT and FOCUSS for high resolution spectrum estimation in our proposed algorithm. It can be concluded from this table that IMAT significantly reduces the simulation time while maintaining acceptable estimation accuracy.

Table 4. Comparison between the AAE and ASTPF Values Achieved for Subject 1 while using IMAT and FOCUSS Spectrum Estimation Algorithms

|  | IMAT | FOCUSS |
|---|---|---|
| Average AAE | 1.72 | 1.65 |
| Average ASTPF | 6.36 | 130.28 |

IV. CONCLUSION

In this paper we proposed a novel algorithm for real-time heart rate estimation using wrist-type PPG signals when subjects are performing intensive physical exercise. In order to deal with the strong MAs caused by subjects' fast running, we recorded and utilized simultaneous acceleration data as the MA reference signals. The proposed algorithm consists of two key steps of MA Cancellation and Spectral Analysis and shows high robustness against strong motion artifacts caused by intensive physical exercise.